\documentclass[english,aps,superscriptaddress,dvipdfmx,reprint,showpacs]{revtex4-1}
\usepackage[T1]{fontenc}
\usepackage[latin9]{inputenc}
\usepackage{verbatim}
\usepackage{bm}
\usepackage{amsmath}
\usepackage{amssymb}
\usepackage{graphicx}
\usepackage{esint}
\usepackage{babel}
\begin{document}

\title{Multipole surface plasmons in metallic nanohole arrays}

\author{Munehiro Nishida}

\email{mnishida@hiroshima-u.ac.jp}

\affiliation{Graduate School of Advanced Science of Matter, Hiroshima University,
Higashi-Hiroshima, 739-8530, Japan}

\author{Noriyuki Hatakenaka}

\affiliation{Graduate School of Integrated Arts and Sciences, Hiroshima University,
Higashi-Hiroshima, 739-8521, Japan}

\author{Yutaka Kadoya}

\affiliation{Graduate School of Advanced Science of Matter, Hiroshima University,
Higashi-Hiroshima, 739-8530, Japan}
\begin{abstract}
The quasi-bound electromagnetic modes for the arrays of nanoholes
perforated in thin gold film are analyzed both numerically by the
rigorous coupled wave analysis (RCWA) method and semi-analytically
by the coupled mode method. It is shown that when the size of the
nanohole occupies large portion of the unit cell, the surface plasmon
polaritons (SPPs) at both sides of the film are combined by the higher
order waveguide modes of the holes to produce \emph{multipole surface
plasmons}: coupled surface plasmon modes with multipole texture on
the electric field distributions. Further, it is revealed that the
multipole texture either enhances or suppresses the couplings between
SPPs depending on their diffraction orders and also causes band inversion
and reconstruction in the coupled SPP band structure. Due to the multipole
nature of the quasi-bound modes, multiple dark modes coexist to produce
variety of Fano resonance structures on the transmission and reflection
spectra.
\end{abstract}


\maketitle

\section{Introduction}

Since the discovery of the extraordinary optical transmission (EOT)
of sub-wavelength hole array systems by Ebbessen \cite{Ebbesen_Wolff_98},
plenty of researches have been devoted to elucidate this mechanism
(for recent review, see \cite{Garcia-Vidal_Kuipers_10}). To date,
the existing theories reveal that this phenomenon comes as a result
of resonances with quasi-bound electromagnetic (EM) modes localized
around the metal film \cite{Lalanne_Hugonin_05, Garcia-Vidal_Kuipers_10}.
Although sub-wavelength holes transmit light
with a very low efficiency in general \cite{Bethe_44,Roberts_87},
the energy of the incident light is carried efficiently to the opposite
side of the film by exciting a quasi-bound mode in the film. Thus
the EOT can be considered as an effect of resonant tunneling of light
via quasi-bound modes.

The quasi-bound modes in metallic nanohole arrays are regarded as
coupled surface plasmon modes, in which the surface plasmon polaritons
(SPPs) at both sides of the metal film are combined by the waveguide
modes in the nanoholes. The contributions of the nanohole array are
two folds. Firstly, the SPP at each side of the film is modified due
to the presence of the holes. This means that the periodic hole array
does not only open band gaps in the SPP band structure but also produce
confinement of the EM fields inside the hole array region. Indeed,
it is known that geometrically induced surface EM modes, also called
as ``spoof surface plasmons'', appear in perforated PECs whose flat
surfaces do not support SPPs \cite{Pendry_04}. Secondly, the (spoof)
surface plasmons at both sides of the film are coupled through the
evanescent fields of the waveguide modes in the holes and form two
separate ``plasmon molecule'' levels \cite{Martin-Moreno_Ebbesen_01}.
In the symmetric environment of equal dielectric constants in the
regions of incidence and transmission, the two molecule levels correspond
to the bonding and anti-bonding modes to which in the (anti-) bonding
mode the SPPs at both sides of the film are combined (anti-) symmetrically
and has a (higher) lower energy.

The propagation constant, $q_{0z}$, of the fundamental waveguide
mode of the nanohole (HE$_{11}$ mode in the case of cylindrical hole)
is the key parameter to determine the property of the coupled SPP
modes. Such that, if $q_{0z}$ becomes nearly zero at the cutoff frequency,
then the confinement of the EM field becomes so large that the dispersions
of the coupled SPP modes flatten. Also, when $q_{0z}=0$, then the
condition corresponds to the zero order Fabry-Perot resonance and
the resulting transmission maxima are much more like cavity resonance
of a single hole \cite{Ruan_Qiu_06,Garcia-Vidal_Martin-Moreno_05}.
Therefore, the coupled SPPs around the cutoff frequency may be considered
as cavity arrays that are weakly coupled by the SPPs, instead. On
the other hand, when the imaginary part of $q_{0z}$ becomes large,
the coupling between the waveguide mode and the SPPs becomes weak
and the coupled SPPs are almost decoupled and behave like (spoof)
surface plasmons on two disconnected metal surfaces. In this case,
the EOT has the same origin as Wood's anomaly \cite{Wood_1902,
Sarrazin_Vigoureux_03}
which is deemed as a result of the EM waves propagating along the metal
surfaces, such as the waves diffracted parallel to the surface of
a PEC (Rayleigh anomaly \cite{Rayleigh_1907}) or by the SPPs propagating
on a real metal surfaces (plasmon anomaly \cite{Hessel_Oliner_65}).

\begin{figure}
\centering{}\includegraphics[width=0.9\columnwidth]{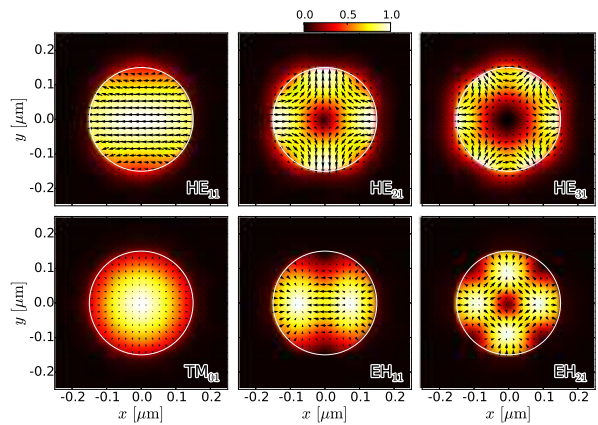}\protect\caption{\label{fig:waveguide_mode}(Color online) Electric field distributions
for the waveguide modes of a nanohole.}
\end{figure}

However, there is another important ingredient of the EOT that has
not been considered much so far. When the size of the hole occupies
large portion of the unit cell, the higher-order waveguide modes would
play an important role. These higher-order waveguide modes as shown
in Figure \ref{fig:waveguide_mode} have multipole nature that weakens
the coupling with incident light that has a dipole nature and hence
leads to having a minor contribution to the transmission spectra.
Nonetheless, it is known that the combination of a bright dipole mode
and a dark multipole mode yields to a sharp peak structure in the
transmission spectra via Fano resonance that focuses much attention
because of the expectation of potential applications such as the high-resolution
bio-sensors \cite{Genet_Woerdman_03, Lukyanchuk_Chong_10}. Therefore, it is expected
that a variety of sharp resonance structures will appear in the transmission
and reflection spectra due to the existence of higher-order waveguide
modes \cite{Liu_Jin_12_with_notes}.

Moreover, it is expected that the multipole modes in the nanoholes
are combined with each other by the SPPs on the film surfaces. In
other words, the SPPs are combined by the multipole modes to produce
novel type of hybridized bound modes which may be called as \emph{multipole
surface plasmons} (MSPs). The band structure of MSPs should have a
distinct feature that is rarely seen in the lattice of dipoles such
as the natural crystal of atoms with higher order of atomic orbitals
that generates various structures of diverse physical properties. 

In this paper, we investigate the contribution of multipolar waveguide
modes of nanoholes to the creation of quasi-bound EM modes for the
arrays of nanoholes perforated in thin gold film. In order to confirm
the existence of MSPs, we have performed a thorough numerical simulation
by using the rigorous coupled wave analysis (RCWA) method, which gives
almost exact results of transmission, reflection and absorption spectra
and also dispersion of quasi-bound modes. The contribution of the
higher-order waveguide modes and their multipolar field distribution
on the band structure of MSPs are further analyzed through the use
of the coupled mode (CM) method.

\section{Method}

We adopt two different theoretical approaches, namely, the RCWA method
\cite{Li_97,Weiss_11} and the CM method \cite{Garcia-Vidal_Kuipers_10,Leon-Perez_Martin-Moreno_08}.
The accuracy of the results obtained from the use of our homemade
numerical codes are checked through comparison with those obtained
by the finite-difference time-domain (FDTD) method \cite{Taflove_Hagness_05}
with the help of a freely available software package \cite{Oskooi_Johnson_10}.
In this section we present the basic ingredients of these two methods
and explain their efficiency in obtaining the dispersion relations
of the quasi-bound modes.

\subsection{RCWA method}

\begin{figure}
\centering{}\includegraphics[width=0.5\columnwidth]{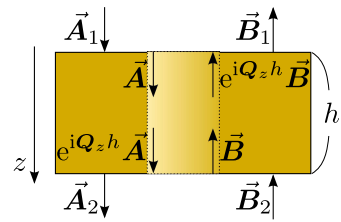}\protect\caption{\label{fig:coefficient_vectors}(Color online) Schematic definition
of the coefficient vectors for the EM modes entering and leaving the
interfaces of layers.}
\end{figure}
The RCWA method is applicable to systems that can be separated effectively
into several planar layers. Each layer is invariant under translation
in the thickness direction which is assumed to be in the $z$-direction
and is periodic in two other spatial directions ($x$- and $y$-directions).
The fields in each layer are expanded in Fourier-Bloch modes that
propagate or decay either forward or backward in the $z$-direction.
We can derive the solution of a stack of layers in a scattering matrix
formalism. Scattering matrix $\mathbb{S}$ provides a relation between
the coefficient vectors $\vec{\bm{A}}_{1}$ and $\vec{\bm{B}}_{2}$
for the input modes and the coefficient vectors $\vec{\bm{A}}_{2}$
and $\vec{\bm{B}}_{1}$ for the output modes as shown below: 
\begin{align}
\begin{bmatrix}\vec{\bm{A}}_{2}\\
\vec{\bm{B}}_{1}
\end{bmatrix} & =\mathbb{S}\begin{bmatrix}\vec{\bm{A}}_{1}\\
\vec{\bm{B}}_{2}
\end{bmatrix},\label{eq:RCWA}
\end{align}
where the vectors $\vec{\bm{A}}_{i}$ ($\vec{\bm{B}}_{i}$) with $i$
being 1 or 2 contain a set of coefficients that define the amplitude
and phase of the EM modes propagating in the $+z$ ($-z$) directions
at the top ($i=1$) or bottom ($i=2$) interfaces (see Fig. \ref{fig:coefficient_vectors}).
In this method, all the EM modes are expressed by the coefficient
vectors with the same number of dimensions $N_{\text{FB}}$ that are
determined by the number of Fourier-Bloch modes used in the expansion.
Likewise, the scattering matrix $\mathbb{S}$ can be constructed iteratively
for the stack of any number of layers considering the propagation
in each layer and the boundary conditions at the interfaces \cite{Weiss_11}.
Conversely, the truncation of Fourier series with a finite value of
$N_{\text{FB}}$ causes serious problem of poor convergence when there
is any discontinuous jump in the permittivity distribution. Although
the convergence can be improved using the so-called Fourier factorization
rules \cite{Li_97}, the improvement is not enough in the case of
metallo-dielectric structures since the difference between the permittivity
of the metallic and the dielectric area is quite large. Therefore,
we adopt the scheme of matched coordinates developed by Thomas Weiss
\cite{Weiss_11}, in which the spatial resolution is increased adaptively
close to the jump discontinuities with the application of the Fourier
factorization rules by introducing a coordinate transformation technique
developed at the beginning of transformation optics \cite{Pendry_Smith_06,Leonhardt_Philbin_10}.
(Figure S1 in the Supplemental Material \cite{SupplementalMaterial}
shows the coordinate we used for the triangular lattice of nanoholes).

From the comparison with the results obtained by the FDTD method,
we see that the RCWA method gives almost exact results for the transmission
and reflection spectra (see Figure S4 in the Supplemental Material).
Throughout our calculations, we used $N_\text{FB}=29\times 29$
Fourier-Bloch modes. Figure S5 in the Supplemental Material shows that
converging behavior is obtained around this number of Fourier-Bloch modes,
and the expected error is small enough.
The computational time required in the RCWA method is significantly
shorter than that required to obtain quantitative results by the FDTD
method.

\subsection{CM method}

In the CM method, the EM fields in the metallic nanohole array region
are expressed only by the waveguide modes of the nanoholes. Imposing
the continuity condition of EM fields at the openings of the holes
and the surface impedance boundary conditions (SIBCs) \cite{Jackson_99}
at the surfaces of the metal film, we can derive a coupled system
of equations for the coefficient vectors as follows:

\begin{align}
 & \mathbb{G}\begin{bmatrix}\vec{\bm{A}}\\
\vec{\bm{B}}
\end{bmatrix}=\begin{bmatrix}\vec{\bm{I}}\\
0
\end{bmatrix},\label{eq:CM}\\
 & \mathbb{G}=\begin{bmatrix}\bm{G}^{-} & \bm{G}^{+}\text{e}^{\text{i}\bm{Q}_{z}h}\\
\bm{G}^{+}\text{e}^{\text{i}\bm{Q}_{z}h} & \bm{G}^{-}
\end{bmatrix},\nonumber 
\end{align}
where $\vec{\bm{A}}$ ($\vec{\bm{B}}$) is the coefficient vector
for the waveguide modes propagating in the $+z$ ($-z$) direction
at the top (bottom) interface. The matrix $\bm{G}^{\pm}$ controls
the EM coupling between waveguide modes at the interfaces, and $\bm{Q}_{z}$
denotes the diagonal matrix whose each diagonal element $q_{\alpha z}$
is the propagation constant associated with mode $\alpha$. The thickness
of the metal film is denoted by $h$ while the vector $\vec{\bm{I}}$
takes into account the direct initial illumination over the waveguide
modes. The mathematical expressions for the different magnitudes can
be found in the Appendix.

At this point, we extend the original formulation \cite{Garcia-Vidal_Kuipers_10}
in two aspects in order to improve its quantitative aspect: (i) waveguide
modes for a real metal with finite permittivity are used for the
calculation of the overlaps of modal wavefunctions with plane waves,
while PEC waveguide modes were approximately used in the original treatment
and (ii) the admittance operator is introduced in order to express
the orthogonality condition for the general hybrid modes having both
TE and TM components. 

Based from the results of the comparison made with the RCWA and FDTD
methods, we have shown that our modifications have improved the CM
method enough to give the quantitative prediction of the whole resonant
structure in the transmission and reflection spectra with a computational
time that is four orders of magnitude shorter than that required by
the FDTD method (see Figure S2-S4 in the Supplemental Material
\cite{SupplementalMaterial}).
Throughout our calculations, we used 91 Fourier-Bloch modes whose
wavenumbers are not exceeding 5 times the magnitude of the
basic reciprocal lattice vector. Figure S6 in the Supplemental Material shows that
converging behavior is obtained around this number of Fourier-Bloch modes,
and the expected error is small enough to obtain quantitative results.

By the mean field approximation for the admittance operator, we can
derive a similar expression as the original CM equations:
\begin{align}
 & \mathbb{G}_{\text{mf}}\begin{bmatrix}\vec{\bm{E}}\\
\vec{\bm{E}}'
\end{bmatrix}=\begin{bmatrix}\vec{\bm{I}}\\
0
\end{bmatrix},\label{eq:CM-mean-field}\\
 & \mathbb{G}_{\text{mf}}=\begin{bmatrix}\bm{G}-\bm{\Sigma} & \bm{G}^{V}\\
\bm{G}^{V} & \bm{G}-\bm{\Sigma}
\end{bmatrix},\nonumber 
\end{align}
where $\vec{\bm{E}}$ ($\vec{\bm{E}}$') approximately represents
the modal amplitude of the electric field at the top (bottom) interface.
The mean field approximation deviates considerably from the exact
calculation in the metal region near the hole edge as we can see from
the in-plane magnetic field distributions shown in the Figure S7 of the
Supplemental Material. However, the overlaps with plane waves are predominantly
determined by the integral in the hole region, the influence on the
spectra is not significant as we can see in the Figure S3 of the
Supplemental Material.

\subsection{Quasi-bound mode}

The occurrence of resonant transmission implies the existence of a
quasi-bound mode in the scattering object. Quasi-bound modes are solutions
of Maxwell's equation that can oscillate in time and hold EM energy
within the object for a considerable period of time in the absence
of incident light. Formally, we would have to derive a non-vanishing
output for zero input.

In the RCWA method, this means that the quasi-bound modes would correspond
to the output vectors $\vec{\bm{A}}_{2}$ and $\vec{\bm{B}}_{1}$
of Eq.\ (\ref{eq:RCWA}) for zero input vectors $\vec{\bm{A}}_{1}=0$
and $\vec{\bm{B}}_{2}=0$. In other words, the vectors $\vec{\bm{A}}_{1}$
and $\vec{\bm{B}}_{2}$ must be enlarged infinitely by the scattering
matrix $\mathbb{S}$. Numerically, the search for quasi-bound modes
can be fulfilled by seeking the infinite singular values of $\mathbb{S}$
using the routine of singular value decomposition.

While in the CM method, the quasi-bound modes would agree to the non-trivial
solutions of Eq.\ (\ref{eq:CM})%
{} for zero input vector $\vec{\bm{I}}=0$, and thus we seek the eigenvectors
of zero eigenvalue (i.e. null space) of the matrix $\mathbb{G}$%
.

In searching for the quasi-bound modes, we must take special care
of the boundary condition at infinity. Quasi-bound modes are not true
bound modes since they have finite lifetime as they eventually lose
energy due to the weak emission in the absence of incident light.
Therefore, we must seek the EM eigenmodes in the lower half of the
complex plane of angular frequency $\omega$, adopting the out-going-wave
boundary condition. This means that one of the branches of the multi-valued
function $k_{z}$ must be chosen obeying the condition that $\text{Re}k_{z}>0$
for $|\text{Re}k_{z}|>|\text{Im}k_{z}|$ or $\text{Im}k_{z}>0$ for
$|\text{Re}k_{z}|<|\text{Im}k_{z}|$. Certainly, this condition becomes
nonphysical such that (i) mode fields explode at infinity (i.e. $\text{Im}k_{z}<0$)
for the modes predominantly propagating to infinity and (ii) energy
flows come from infinity (i.e. $\text{Re}k_{z}<0$) for the predominantly
evanescent modes \cite{Taylor_72}.

\section{Results}

\subsection{Transmission, Reflection and Absorption Spectra}

\begin{figure}
\centering{}\includegraphics[width=0.8\columnwidth]{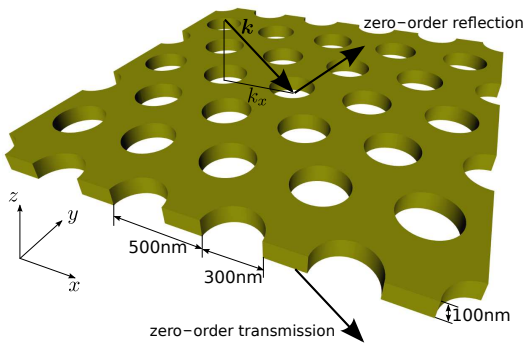}\protect\caption{\label{fig:system}(Color online) Schematic diagram of a triangular
lattice of nanoholes perforated in a thin gold film.}
\end{figure}

The system that we are concerned in this paper is a triangular lattice
of nanoholes perforated in a thin gold film with a thickness of $h=100$nm
as shown in Figure \ref{fig:system}. The film is surrounded with
water. The diameter of the hole is assumed to be 300nm and the period
of array, $d$, is 500nm. The incident light is assumed to be a p-wave,
and the component of the wave vector parallel with the film surface
is in the $x$ direction, which is parallel to one of the lattice vectors. 

\begin{figure}
\begin{centering}
\includegraphics[width=1\columnwidth]{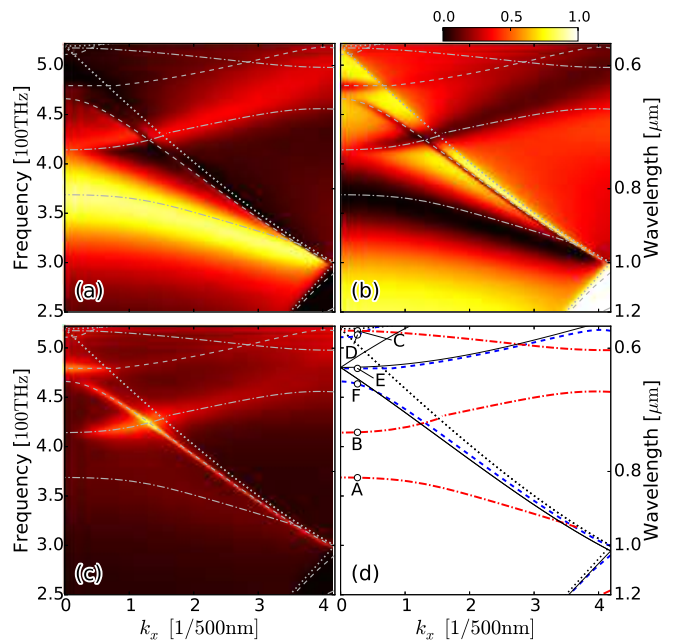}
\par\end{centering}

\centering{}\protect\caption{\label{fig:spectra}(Color online) Zero-order transmission spectra
(a), zero-order reflection spectra (b), and total absorption spectra
(c) for a triangular lattice of nanohole array. Gray dashed and dash-dotted
lines indicate the dispersion of bound surface plasmons (BSPs) and
gray dotted lines indicate the condition of the Rayleigh anomaly.
The bonding (anti-bonding) BSPs are shown by the red dash-dotted (blue
 dashed) lines in (d) together with the black solid lines for the
 empty-lattice bands.}
\end{figure}

Figure \ref{fig:spectra} shows the spectra of (a) zero-order transmission,
(b) zero-order reflection and (c) total absorption. These results
were obtained by numerical simulation based on the RCWA method. The
panel (d) shows the dispersion relation of the quasi-bound modes (red
dash-dotted and blue dashed lines) together with the Rayleigh anomaly
(dotted lines) and the dispersions of SPPs estimated using empty-lattice
approximation (black solid lines). The dispersion of the quasi-bound
modes and the Rayleigh anomaly are also shown in the panels (a)-(c)
by gray dashed (or dash-dotted) lines and gray dotted lines respectively. At this point,
we can clearly see the coincidence between resonant peaks or dips
in the spectrum and the dispersions of quasi-bound modes. Although
there is an abrupt change on the line of the Rayleigh anomaly, the
profile of this change is still quite different. Hence, we conclude
that the resonant structure of spectra is produced by quasi-bound
modes.

Furthermore, we can also see that there are crossing points between
red and blue lines. This means that there are two kinds of
modes which cannot couple with each other due to the difference in
symmetry. Since these two modes contribute independently to the absorption,
it is expected that the total absorption would become high at the
crossing points. Indeed, at the crossing point between the blue line
passing through the point F and the red line passing though the point
B, the total absorption rate becomes quite high, to about more than
90\% as shown in the panel (c).

\subsection{Bonding and Anti-bonding Bound Modes}

\begin{figure}
\centering{}\includegraphics[width=0.7\columnwidth]{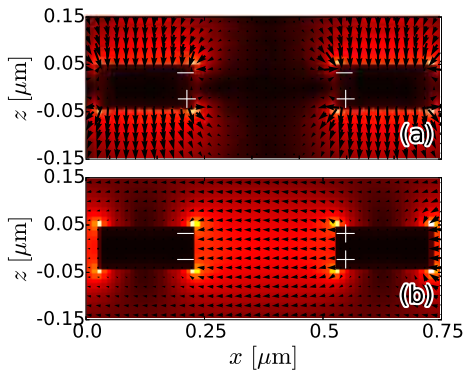}\protect\caption{\label{fig:efield_xz}(Color online) Electric field distributions
in the $xz$ plane for an anti-bonding mode (a), and a bonding mode
(b).}
\end{figure}

As we pointed out in the last subsection, there are two types of branches
as represented by the red dash-dotted lines passing through points
A, B and C and blue dashed lines passing through points D, E and F
in the panel (d) of Figure \ref{fig:spectra}. The blue dashed lines
are almost in accordance with the black solid lines, which indicate
the empty-lattice bands of SPPs, while the slope of the red dash-dotted
lines are rather shallow.

\begin{figure}
\centering{}\includegraphics[width=1\columnwidth]{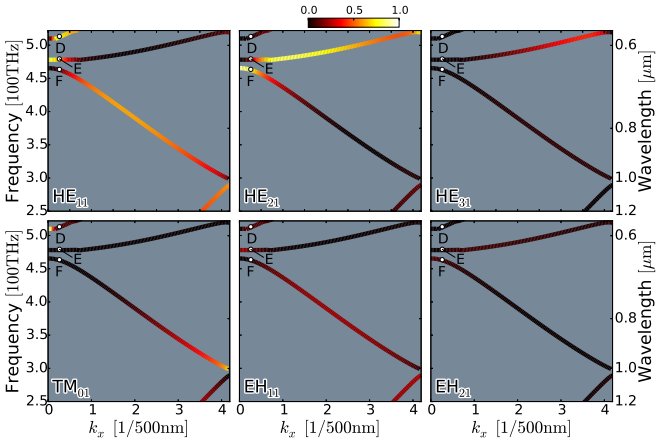}\protect\caption{\label{fig:anti_bonding_bound_mode_contribution}(Color online) Rate
of contribution of each waveguide mode to create anti-bonding bound
modes calculated using the coupled-mode method.}
\end{figure}

Figure \ref{fig:efield_xz} presents the electric field distributions
in the $xz$ plane (a) at point F and (b) at point A as shown in the
panel (d) of Figure \ref{fig:spectra}. From the panel (a), we see
that the charge distribution is anti-symmetric about the $z=0$ plane,
and the intensity of the electric field strengthens mainly outside
of the metal surfaces between the holes. On the other hand, from the
panel (b), we see that the charge distribution is symmetric about
the $z=0$ plane and the intensity of the electric field concentrates
around the holes, specifically inside of the metal film region. These
features indicate that the modes on the blue lines are anti-bonding
(AB) modes in which the SPPs of both sides of the gold film are combined
anti-symmetrically like a long-range SPP of a thin metal film \cite{Sarid_81}.
On the other hand, the modes on the red lines are bonding (B) modes
like a short-range SPP of a thin metal film. The symmetry of each
bound mode can be checked clearly using the CM method. The dispersion
relations of B-SPP and AB-SPP calculated using the CM method (Eq.\ (\ref{eq:CM}))
under the symmetric ($\vec{\bm{A}}=\vec{\bm{B}})$ and anti-symmetric
($\vec{\bm{A}}=-\vec{\bm{B}}$) conditions are shown in Figures \ref{fig:bonding_bound_mode_contribution}
and \ref{fig:anti_bonding_bound_mode_contribution}, respectively.

In the AB-SP where the electric field is expelled from the hole area,
each SPP propagates on each surface without feeling much of the holes.
This in turn makes the dispersions sit near the empty-lattice bands.
We can also see that the resonant structure of the spectra created
by AB-SPP is narrow due to the darkness of the mode. On the other
hand, the B-SPP has large contribution from the waveguide modes in
the holes and the dispersions are shifted largely from the empty-lattice
bands. The shallowness of the band indicates that the B-SPP is much
like a cavity mode whose electric field is confined in the nanoholes.

The lowest energy B-SPP mode can also be considered as a coupled spoof
surface plasmon, since the dispersion saturates around the cutoff
frequency of the HE$_{11}$ mode, 355 THz. But then what are higher
energy B-SPP modes? These modes may be considered as \emph{coupled
multipole spoof surface plasmons} which are geometrically induced
coupled surface modes created by the coupling between the SPP on the
metal surface and the higher order waveguide modes of the nanoholes
as discussed in the next subsection.

\subsection{Multipole Nature of Bound Modes}

\begin{figure}
\centering{}\includegraphics[width=1\columnwidth]{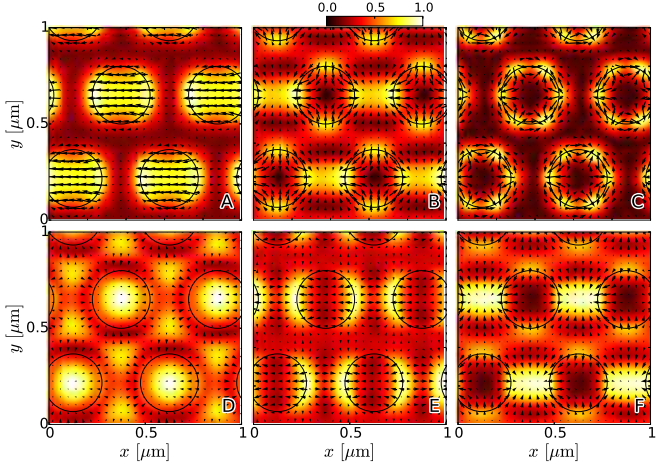}\protect\caption{\label{fig:Electric-field-distributions}(Color online) Electric field
distributions at 20nm above the surface of the gold film for the points
shown in Fig. \ref{fig:spectra}(d).}
\end{figure}

Figure \ref{fig:Electric-field-distributions} shows the electric
field distributions at 20nm above the surface of the gold film for
the quasi-bound modes at the points shown in Figure \ref{fig:spectra}(d).
We can see monopole (D), dipole (A, E), quadrupole (B, F) or hexapole
(C) texture for each branch of the quasi-bound modes. The most striking
feature is that a strong coupling between adjacent holes seems to
exist in the quadrupole texture (B, F) as well as positive charges
are aligned in the $x$-direction to form a stripe texture.

\begin{figure}
\centering{}\includegraphics[width=1\columnwidth]{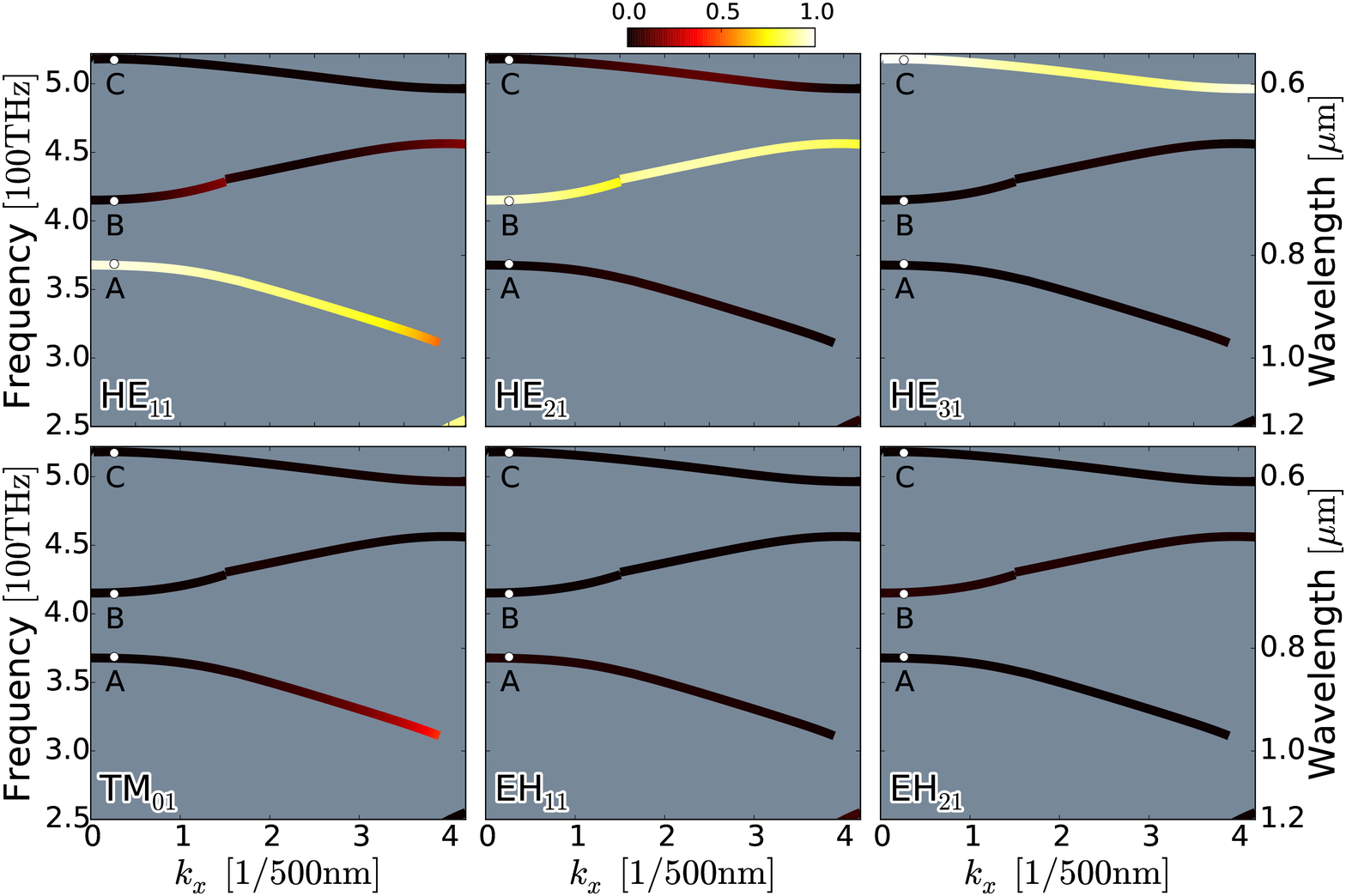}\protect\caption{\label{fig:bonding_bound_mode_contribution}(Color online) Rate of
contribution of each waveguide mode to create bonding bound modes
calculated using the coupled-mode method.}
\end{figure}

In order to see the role played by the waveguide modes to create these
multipole textures, we have analyzed the contribution of waveguide
modes for each branch using the CM method. Figure \ref{fig:bonding_bound_mode_contribution}
shows the rate of contribution of the waveguide mode whose name is
written at the bottom left corner of each panel. The bright color
means that the rate of contribution of the mode is high. It is clearly
seen that each branch of the B-SPPs namely, A, B and C is mainly created
by a single waveguide mode, HE$_{11}$, HE$_{21}$ or HE$_{31}$ respectively.
This means that the coupled spoof surface plasmon yields higher energy
branches with multipole textures according to the higher order waveguide
modes.

The contribution of the waveguide modes to the AB-SPPs is more complicated.
There is non-negligible contribution from the TM-like modes, i.e.
TM$_{01}$ and EH$_{11}$. Furthermore, a kind of band inversion seems
to occur between two bands produced by HE$_{11}$ and HE$_{21}$ modes.
The contribution rates of HE$_{11}$ and HE$_{21}$ modes are reversed
around the points E and F. This could be explained as a result of
band anti-crossing arising from the band inversion between HE$_{11}$
and HE$_{21}$ bands near the $\Gamma$ point.

In order to reveal the origin of these band structures, we approximately
analyze the dispersion relation of the quasi-bound mode produced by
a single waveguide mode either from HE$_{11}$ or HE$_{21}$. This
is done using a metamaterial treatment as in \cite{Garcia-Vidal_Kuipers_10},
where only a single mode $\alpha(=\text{HE}_{11}\text{ or HE}_{21})$
inside the hole is taken into account and only the p-polarized zero-order
diffraction mode is considered in the quantity $G_{\alpha\alpha}$
governing the coupling between holes. Starting with the mean-field
approximated CM equations (\ref{eq:CM-mean-field}), we obtain the
condition for the existence of B-SPP mode as $\left(G_{\alpha\alpha}+G_{\alpha}^{V}-\Sigma_{\alpha}\right)E_{\alpha}=0$
where $G_{\alpha\alpha}=\left|S\right|^{2}Y_{\vec{k}\text{p}}/f_{\vec{k}\text{p}}^{+}$,
$S=\left\langle \alpha\left|\vec{k}\text{p}\right.\right\rangle $,
$\Sigma_{\alpha}-G_{\alpha}^{V}=Y_{\alpha}/\left(Z_{\text{s}}Y_{\alpha}-\gamma\right)$
and $\gamma=\text{i}/\tan\left(q_{\alpha z}h/2\right)$. The definitions
of the symbols can be found in the Appendix. Neglecting the imaginary
part of the permittivity of gold $\varepsilon_{\text{m}}$, the dispersion
relation of the bonding quasi-bound mode is given by 
\begin{equation}
k_{x}=\sqrt{\varepsilon}k_{\omega}\sqrt{1+\frac{\varepsilon}{\left|\varepsilon_{\text{m}}\right|}\left\{ \left(1-|S|^{2}\right)+\frac{|S|^{2}\left|\gamma\sqrt{\varepsilon_{\text{m}}}\right|}{Z_{0}\left|Y_{\alpha}\right|}\right\} ^{2}}.\label{eq:bonding_spoof_dispersion}
\end{equation}

From Eq.\ (\ref{eq:bonding_spoof_dispersion}), we can see that the
dispersion relation is reduced to that of the SPP on the flat gold
surface within the SIBC approach, $k_{x}\simeq\sqrt{\varepsilon}k_{\omega}\sqrt{1-\varepsilon/\varepsilon_{\text{m}}}\simeq k_{\omega}\sqrt{\frac{\varepsilon\varepsilon_{\text{m}}}{\varepsilon+\varepsilon_{\text{m}}}}$,
if the coupling between the waveguide mode and the diffraction mode
is negligibly small $(|S|^{2}\simeq0)$. On the other hand, as the
frequency approaches the cutoff frequency from below, $|\gamma/Y_{\alpha}|$
diverges for HE modes since HE modes approach TE modes and the effective
admittance $Y_{\alpha}$ goes to zero as $q_{\alpha z}$ goes to zero.
Correspondingly, $\gamma$ goes to $\infty$ in this limit. Although
the absorption of gold prevents from diverging, this divergence property
remains and the dispersion relation flattens around the cutoff frequency
as long as the film remains thin.

\begin{figure}
\centering{}\includegraphics[width=1\columnwidth]{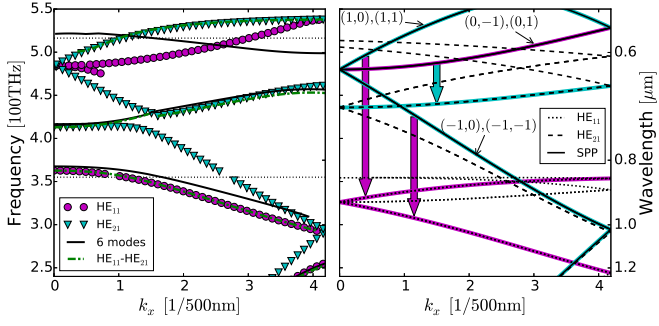}\protect\caption{\label{fig:bonding_mode_dispersions}(Color online) Dispersion relations
of the bonding quasi-bound modes produced by each single waveguide
mode. The left panel shows those of HE$_{11}$ (magenta circles) and
HE$_{21}$ (cyan triangles) modes. The black solid lines and the green
dash-dotted lines indicate those of 6 modes and two modes. The right
panel shows those of HE$_{11}$ (dotted line) and HE$_{21}$ (dashed
line) modes obtained approximately with metamaterial treatments. The
black solid lines indicate empty-lattice bands of the SPP on a flat
gold surface. The magenta (cyan) solid lines indicate the branches
selected to form the quasi-bound modes of HE$_{11}$ (HE$_{21}$)
as a guide for the eye.}
\end{figure}

The right panel of Figure \ref{fig:bonding_mode_dispersions} shows
the dispersion relation of Eq.\ (\ref{eq:bonding_spoof_dispersion})
considering the diffraction effects in the empty-lattice limit. We
can see that the dispersion curves, dotted lines and dashed lines,
converge to the cutoff frequencies of HE$_{11}$ mode (355THz) and
HE$_{21}$ mode (516THz), respectively. %
In this panel, the empty-lattice bands of the SPPs on a flat gold
surface are also shown by black solid lines. The set of numbers $(m,n)$
indicates the diffraction order which mainly contributes to the branch
pointed by the arrow. The parallel momentum $\vec{k}$ for the diffraction
order $(m,n)$ is described by the expression $\vec{k}=\left(k_{x},0\right)+m\vec{b}_{1}+n\vec{b}_{2}$
where $\vec{b}_{1}$ and $\vec{b}_{2}$ are the reciprocal lattice
vectors for the triangular lattice of the period $d$ given by $\vec{b}_{1}=\frac{2\pi}{d}\left(1,-\frac{1}{\sqrt{3}}\right),\quad\vec{b}_{2}=\frac{2\pi}{d}\left(0,\frac{2}{\sqrt{3}}\right)$.
Therefore, the branches for the diffraction orders $(1,0)$, $(1,1)$,
$(-1,0)$ and $(-1,-1)$ are composed of the EM wave with a period
equal to $\sqrt{3}d$, such that
\begin{align*}
 & \text{e}^{\text{i}\left\{ \left(k_{x}\pm\frac{2\pi}{d}\right)x+\frac{2\pi}{\sqrt{3}d}y\right\} }+\text{e}^{\text{i}\left\{ \left(k_{x}\pm\frac{2\pi}{d}\right)x-\frac{2\pi}{\sqrt{3}d}y\right\} }\\
 & =2\text{e}^{\text{i}\left(k_{x}\pm\frac{2\pi}{d}\right)x}\cos\left(\frac{2\pi}{\sqrt{3}d}y\right).
\end{align*}
Also, the branch for the diffraction orders $(0,-1)$ and $(0,1)$
is composed of the EM wave with a period equivalent to $\sqrt{3}d$/2,
where
\[
\text{e}^{\text{i}\left\{ k_{x}x-\frac{4\pi}{\sqrt{3}d}y\right\} }+\text{e}^{\text{i}\left\{ k_{x}x+\frac{4\pi}{\sqrt{3}d}y\right\} }=2\text{e}^{\text{i}k_{x}x}\cos\left(\frac{4\pi}{\sqrt{3}d}y\right).
\]

These periodicities in the $y$-direction are considered to play a
major role to produce the bands of B-SPPs. The left panel of Figure
\ref{fig:bonding_mode_dispersions} shows the dispersion relations
calculated by the mean-field approximated CM equations (\ref{eq:CM-mean-field})
including only a single waveguide mode either HE$_{11}$ (magenta
circles) or HE$_{21}$ (cyan triangles) but with the presence of as
many diffraction modes as needed to achieve convergence. Taking into
account the fact that the dispersion relation moves closer to the
light line if higher-order diffraction modes are included \cite{DeAbajo_Saenz_05,Hendry_Sambles_08},
it is reasonable to think that the branches of B-SPP bands created
by HE$_{11}$ mode are composed of two types of branches: (i) the
$\sqrt{3}d$ periodic branches from the flat-surface SPP bands that
correspond to the black solid lines in the right panel and (ii) the
$\sqrt{3}d$/2 periodic branches from the bands of coupled spoof surface
plasmons that represent the dotted lines in the right panel. The diffracted
waves with $\sqrt{3}d$ periodicity may couple strongly with the dipole
lattices created by HE$_{11}$ mode. In effect, coupled spoof surface
plasmons are formed and dispersion relation are put downward since
positive and negative charges align alternately in the $y$-direction
to form $\sqrt{3}d$ periodic texture as you can see in Figure \ref{fig:Electric-field-distributions}
A. Besides, the diffracted waves with $\sqrt{3}d$/2 periodicity may
couple strongly with the quadrupole lattices created by HE$_{21}$
mode since positive or negative charges produce bond between adjacent
holes to form stripe texture with $\sqrt{3}d$/2 periodicity as you
can see in Figure \ref{fig:Electric-field-distributions} B. Considering
the couplings between HE$_{11}$ and HE$_{21}$ or with higher order
waveguide modes, the texture seems to be optimized and yield simpler
band structure (see green dash-dotted lines and black solid lines
in the left panel). 

\begin{figure}
\centering{}\includegraphics[width=1\columnwidth]{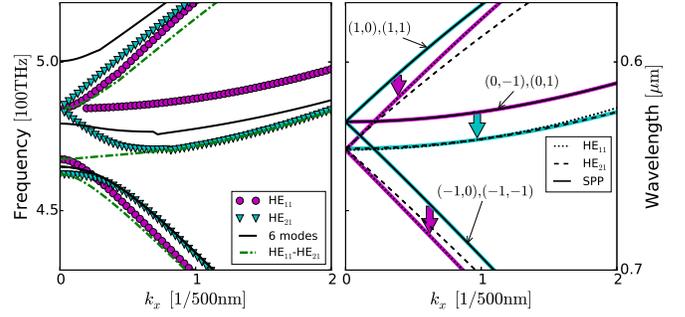}\protect\caption{\label{fig:anti_bonding_mode_dispersions}(Color online) Dispersion
relations of the anti-bonding quasi-bound modes produced by a single
waveguide mode. The meaning of each line is the same as in Figure
\ref{fig:bonding_mode_dispersions}. The lines at the right panel
correspond to a thicker film with thickness of 150nm.}
\end{figure}

The dispersion relation of the AB-SPP mode is obtained by replacing
$\gamma$ with $\gamma^{-1}$ in Eq.\ (\ref{eq:bonding_spoof_dispersion})
as shown in the right panel of Figure \ref{fig:anti_bonding_mode_dispersions}.
In this case, the influence of the coupling between waveguide mode
and diffraction modes become weak as $h$ becomes small. The results
for the film with $h=150$nm are shown in figure for clarity. The
left panel shows the dispersion relations of AB-SPP obtained with
the same procedure as used in the left panel of Figure \ref{fig:bonding_mode_dispersions}.
These results are seemingly produced by the same mechanism as those
of B-SPP: only the $\sqrt{3}d$ ($\sqrt{3}d$/2) periodic branches
are put downward due to the periodicity of the dipole (quadrupole)
texture for the HE$_{11}$ (HE$_{21}$) mode. Therefore, the multipole
textures are the origin of the band inversion and anti-crossing between
HE$_{11}$ and HE$_{21}$ bands.

\subsection{Fano Resonance}

\begin{figure}
\centering{}\includegraphics[width=0.48\columnwidth]{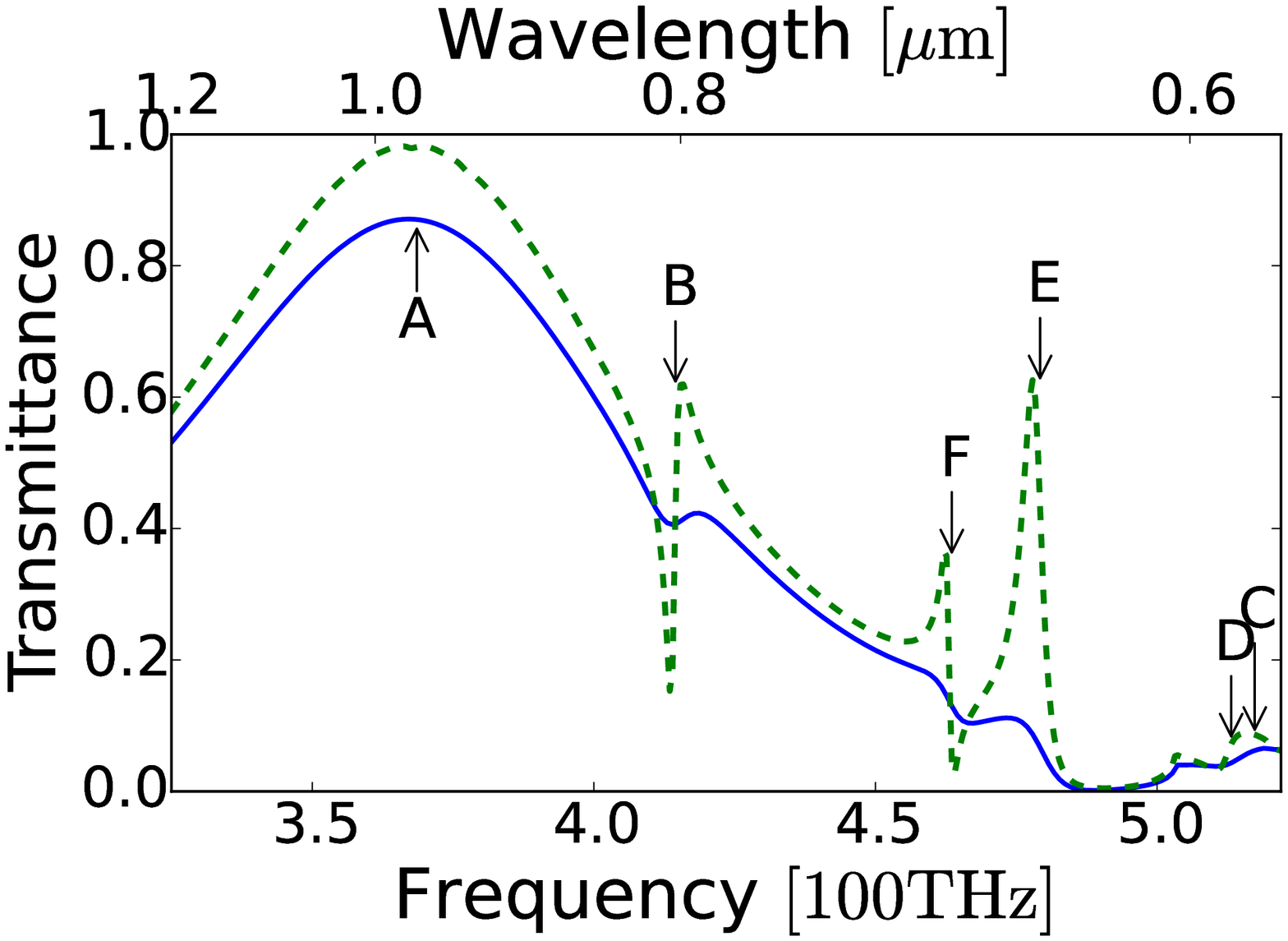}
\includegraphics[width=0.48\columnwidth]{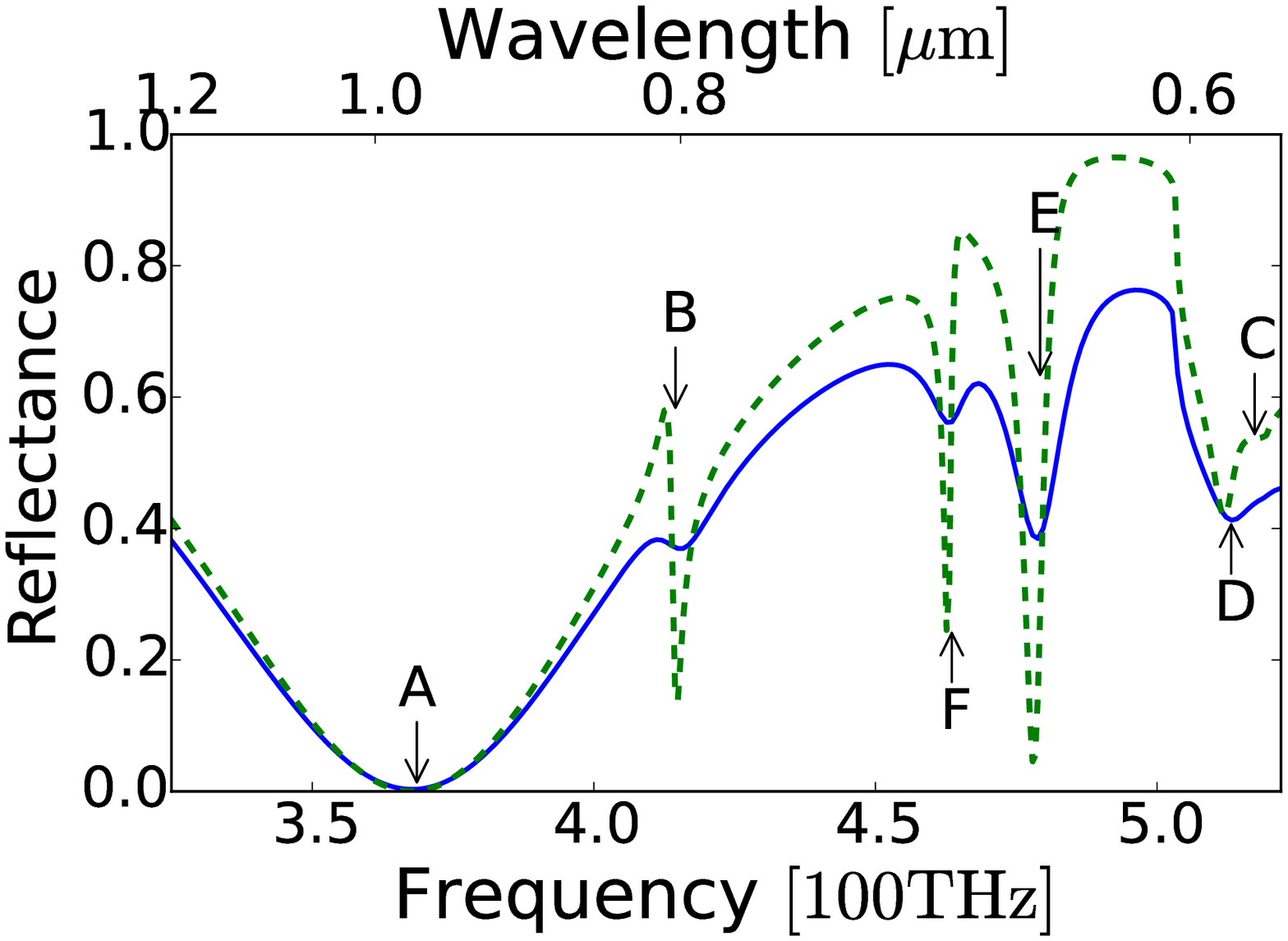}\protect\caption{\label{fig:Fano}(Color online) Fano structures in the transmission
(left panel) and reflection (right panel) spectra.}
\end{figure}

As mentioned above, the AB-SPPs and the higher energy B-SPPs have
multipole nature, which can be considered as dark modes. The peak
and dip structures in the spectra would be attributed to the Fano
resonances between the brighter modes and the darker modes. Indeed,
if the loss of gold is low to about 10\% of the normal loss, then
we can clearly see a typical asymmetric Fano structures that consist
of pairs of peak and dip at nearby frequencies \cite{Genet_Woerdman_03,
Lukyanchuk_Chong_10} as shown in Figure \ref{fig:Fano}.

Thus, the complicated and distinct structure in the transmission and
reflection spectra of this system can be attributed to the Fano resonances
between various orders of multipole surface plasmons.

\section{Conclusion}

By thorough numerical simulations using RCWA and CM methods, we have
found that the higher order waveguide modes of nanoholes combine the
two SPPs at both sides of metal film to create two types of quasi-bound
EM modes. These are the (i) AB-SPP with anti-symmetric charge distribution
and (ii) B-SPP with symmetric charge distribution about the center
plane of the film. Due to the multipole nature of the waveguide modes,
the electric field distribution near the film surface forms a multipole
texture. The Fano resonances between various orders of these multipole
surface plasmons produce sharp peak-dip structure in the transmission
and reflection spectra. 

The B-SPP modes can be considered as coupled multipole spoof surface
plasmons whose electric fields are mainly confined in the nanohole
region and the waveguide modes play a major role in determining their
dispersions. In other words, the nanoholes act like artificial atoms
and the multipolar EM wave in a hole hops to other hole via SPPs just
like the electron in a lattice of real atoms that have higher atomic
orbitals. More so, the dispersion of the AB-SPP mode is determined
by the combination of multiple waveguide modes which yields a fine
band structure due to the band inversion and anti-crossing induced
by the multipole textures.

Further study of these two types of multipole surface plasmons will
open the possibility to create various artificial materials with high
functionality utilizing their high degrees of freedom, e.g. a perfect
absorber using multiple crossing points of the band structure or a
high-resolution bio-sensor using multiple resonant transmissions.

\appendix

\section{Coupled-Mode Method for Real Metal}

In general, the mode fields of a waveguide that is composed of a piecewise
homogeneous media can be expressed as a linear combination of TE and
TM components, such that
\begin{align}
\vec{E}_{\alpha}(\vec{r}) & =a_{\alpha}(\vec{r})\vec{E}_{\alpha\text{TE}}(\vec{r})+b_{\alpha}(\vec{r})\vec{E}_{\alpha\text{TM}}(\vec{r}),\\
-\bm{e}_{z}\times\vec{H}_{\alpha}(\vec{r}) & =a_{\alpha}(\vec{r})Y_{\alpha\text{TE}}(\vec{r})\vec{E}_{\alpha\text{TE}}(\vec{r})\nonumber \\
 & +b_{\alpha}(\vec{r})Y_{\alpha\text{TM}}(\vec{r})\vec{E}_{\alpha\text{TM}}(\vec{r}),
\end{align}
where $\vec{E}_{\alpha}$ and $\vec{H}_{\alpha}$ are the electric
and magnetic field vectors that are respectively projected onto the
$xy$-plane. Also, $\bm{e}_{z}$ is a unit vector along the $z$-axis
and $\vec{r}=(x,$y) is the position vector in the $xy$-plane. Here,
the mode index $\alpha$ represents the full information of the waveguide
mode of a nanohole in the metal film region, such as the ``HE$_{11}$
horizontal mode'' \cite{Roberts_87}. It may also represent the parallel
momentum $\vec{k}$ and the polarization $\sigma(=\text{p or s)}$
for plane-wave modes in the homogeneous dielectric layers where $\left(a_{\vec{k}\text{s}},b_{\vec{k}\text{s}}\right)=(1,$0)
or $\left(a_{\vec{k}\text{p}},b_{\vec{k}\text{p}}\right)=(0,1)$.
The position-dependent admittances for the TE and TM modes are piecewise
constants whose values in the $i$th homogeneous medium are given
by
\begin{align}
Y_{\alpha\text{TE}}^{(i)} & =\frac{1}{\mu_{i}Z_{0}}\frac{q_{\alpha z}}{k_{\omega}},\\
Y_{\alpha\text{TM}}^{(i)} & =\frac{\varepsilon_{i}}{Z_{0}}\frac{k_{\omega}}{q_{\alpha z}},
\end{align}
where $Z_{0}$ and $k_{\omega}$ are the impedance and the wavenumber
in the vacuum, $\varepsilon_{i}$ and $\mu_{i}$ are the relative
permittivity and relative permeability in the $i$th homogeneous medium,
and $q_{\alpha z}$ is the propagation constant of the mode $\alpha$.
This representation indicates that the magnetic field of a mode is
constructed by a position-dependent admittance from the corresponding
electric field. If we use Dirac's notation to describe the electric
field, such that 
\begin{equation}
\vec{E}_{\alpha}(\vec{r})=\langle\vec{r}|\alpha\rangle,
\end{equation}
the effect of the position-dependent admittance is expressed by the
admittance operator $\hat{Y}$ which maps the electric field to the
corresponding magnetic field, such that

\begin{align}
-\bm{e}_{z}\times\vec{H}_{\alpha}(\vec{r}) & =\left\langle \vec{r}\left|\hat{Y}\right|\alpha\right\rangle .
\end{align}
 The orthogonality condition for the modes can be expressed by the
admittance operator as 
\begin{align}
\left\langle \alpha\left|\hat{Y}\right|\beta\right\rangle  & =\int_{-\infty}^{\infty}\text{d}x\int_{-\infty}^{\infty}\text{d}y\vec{E}_{\alpha}^{*}\cdot\left(-\bm{e}_{z}\times\vec{H}_{\beta}\right)\nonumber \\
 & =\int_{-\infty}^{\infty}\text{d}x\int_{-\infty}^{\infty}\text{d}y\vec{E}_{\alpha}^{*}\times\vec{H}_{\beta}=Y_{\alpha}\delta_{\alpha\beta},\\
Y_{\alpha} & =\langle\alpha|\hat{Y}|\alpha\rangle
\end{align}
where $\delta_{\alpha\beta}$ is the Kronecker delta and $*$ denotes
the complex conjugate. 

Using this definition of the admittance operator, the CM equation
(\ref{eq:CM}) can be derived in a similar manner as the original
derivation \cite{Leon-Perez_Martin-Moreno_08}. The $\alpha$ component
of the coefficient vector $\vec{\bm{I}}$ and the $\alpha\beta$ component
of the matrix $\bm{G}^{\pm}$ that appeared in Eq.\ \ref{eq:CM}
are expressed as

\begin{align}
I_{\alpha} & =\frac{2}{f_{\vec{k}\sigma}^{+}}\left\langle \alpha\left|\vec{k}_{0}\sigma_{0}\right.\right\rangle ,\\
G_{\alpha\beta}^{\pm} & =\sum_{\vec{k}\sigma}\frac{Y_{\vec{k}\sigma}}{f_{\vec{k}\sigma}^{+}}\left\langle \alpha\left|\vec{k}\sigma\right.\right\rangle \left\langle \vec{k}\sigma\left|\hat{f}^{\pm}\right|\beta\right\rangle \mp Y_{\alpha}\delta_{\alpha\beta}.
\end{align}
Here, we introduced an operator $\hat{f}^{\pm}=1\pm Z_{\text{s}}\hat{Y}$
which has an expectation value $f_{\vec{k}\sigma}^{\pm}=1\pm Z_{\text{s}}Y_{\vec{k}\sigma}$
for plane waves using the surface impedance 
\begin{equation}
Z_{\text{s}}=\sqrt{\frac{\mu_{\text{m}}}{\varepsilon_{\text{m}}}}Z_{0}
\end{equation}
with $\varepsilon_{\text{m}}$ and $\mu_{\text{m}}$ being the relative
permittivity and the relative permeability of the metal. Additionally,
the transmission and reflection coefficients are expressed as
\begin{align}
t_{\vec{k}\sigma} & =\frac{1}{f_{\vec{k}\sigma}^{+}}\sum_{\alpha}\left\{ \left\langle \vec{k}\sigma\left|\hat{f}^{+}\right|\alpha\right\rangle A_{\alpha}\text{e}^{\text{i}q_{\alpha z}h}+\left\langle \vec{k}\sigma\left|\hat{f}^{-}\right|\alpha\right\rangle B_{\alpha}\right\} ,\\
r_{\vec{k}\sigma} & =-\frac{f_{\vec{k}\sigma}^{-}}{f_{\vec{k}\sigma}^{+}}\delta_{\vec{k}\vec{k}_{0}}\delta_{\sigma\sigma_{0}}\nonumber \\
 & +\frac{1}{f_{\vec{k}\sigma}^{+}}\sum_{\alpha}\left\{ \left\langle \vec{k}\sigma\left|\hat{f}^{-}\right|\alpha\right\rangle A_{\alpha}+\left\langle \vec{k}\sigma\left|\hat{f}^{+}\right|\alpha\right\rangle \text{e}^{\text{i}q_{\alpha z}h}B_{\alpha}\right\} ,
\end{align}
where $A_{\alpha}$ ($B_{\alpha}$) is the coefficient for the mode
$\alpha$ propagating in the $+z$ ($-z$) direction at the top (bottom)
interface, and $h$ denotes the film thickness.

If we adopt a kind of mean-field approximation for the admittance
operator, such that
\begin{align}
\left\langle \vec{k}\sigma\left|\hat{Y}\right|\alpha\right\rangle  & \simeq Y_{\alpha}\left\langle \left.\vec{k}\sigma\right|\alpha\right\rangle ,\\
\left\langle \vec{k}\sigma\left|\hat{f}^{\pm}\right|\alpha\right\rangle  & \simeq\left(1\pm Z_{\text{s}}Y_{\alpha}\right)\left\langle \left.\vec{k}\sigma\right|\alpha\right\rangle \equiv f_{\alpha}^{\pm}\left\langle \left.\vec{k}\sigma\right|\alpha\right\rangle ,
\end{align}
and introduce the modal amplitudes of the electric field at the top
and bottom interfaces as
\begin{align}
E_{\alpha} & =A_{\alpha}f_{\alpha}^{-}+B_{\alpha}\text{e}^{\text{i}q_{\alpha z}h}f_{\alpha}^{+},\\
E_{\alpha}' & =A_{\alpha}\text{e}^{\text{i}q_{\alpha z}h}f_{\alpha}^{+}+B_{\alpha}f_{\alpha}^{-},
\end{align}
then the CM equations can be expressed as Eq.\ (\ref{eq:CM-mean-field}).
The $\alpha\beta$ component of the matrices $\bm{G}$, $\bm{\Sigma}$
and $\bm{G}^{V}$ are given as
\begin{align}
G_{\alpha\beta} & =\sum_{\vec{k}\sigma}\frac{Y_{\vec{k}\sigma}}{f_{\vec{k}\sigma}^{+}}\left\langle \alpha\left|\vec{k}\sigma\right.\right\rangle \left\langle \left.\vec{k}\sigma\right|\beta\right\rangle ,\\
\Sigma_{\alpha\beta} & =\delta_{\alpha\beta}Y_{\alpha}\frac{f_{\alpha}^{+}\text{e}^{\text{i}q_{\alpha z}h}+f_{\alpha}^{-}\text{e}^{-\text{i}q_{\alpha z}h}}{\left(f_{\alpha}^{+}\right)^{2}\text{e}^{\text{i}q_{\alpha z}h}-\left(f_{\alpha}^{-}\right)^{2}\text{e}^{-\text{i}q_{\alpha z}h}},\\
G_{\alpha\beta}^{V} & =\delta_{\alpha\beta}\frac{2Y_{\alpha}}{\left(f_{\alpha}^{+}\right)^{2}\text{e}^{\text{i}q_{\alpha z}h}-\left(f_{\alpha}^{-}\right)^{2}\text{e}^{-\text{i}q_{\alpha z}h}}.
\end{align}

The details of the derivation will be published in another avenue.
\begin{acknowledgments}
This work was supported in part by KAKENHI No. 23651149 from MEXT
of Japan.
\end{acknowledgments}

%

\end{document}